\documentclass[a4paper,11pt]{article}
\pdfoutput=1 

\usepackage{jheppub} 

\usepackage[T1]{fontenc} 

\title{\boldmath Effect of Coulomb energy on Skyrmions}

\author[a]{Nana Ma}
\author[b]{Chris James Halcrow}
\author[a,1]{Hongfei Zhang\note{Corresponding author.}}

\affiliation[a]{School of Nuclear Science and Technology, Lanzhou University,\\ 730000 Lanzhou, P.R. China}
\affiliation[b]{School of Mathematics, University of Leeds, \\Leeds, LS2 9JT, UK}

\emailAdd{mann15@lzu.edu.cn}
\emailAdd{C.J.Halcrow@leeds.ac.uk}
\emailAdd{zhanghongfei@lzu.edu.cn}

\abstract{The Coulomb effect, an essential ingredient in nuclear systems, is quantitatively investigated for certain Skyrmions within the Skyrme model. To do this we calculate the Coulomb energy from numerically generated multi-Skyrmions and, simultaneously, introduce an effective alpha-like particle approximation for large Skyrmions with baryon number $B$, where $B$ is a multiple of four. The calculated Coulomb energies and the corresponding fitted curve from this approximation match well with the results from the numerical method, as well as the behavior of the Coulomb energy in the semi-empirical mass formula. The Skyrme model correctly reproduces the well known results for the Coulomb energy of nuclei. This suggests that the alpha-particles can be used as the fundamental degrees of freedom in the Skyrme model. The Coulomb effect on the binding energy of the Skyrmions increases with baryon number but has a small effect overall. However, the effect could be significantly more pronounced for loosely bound Skyrme models.}

\begin{document}
\maketitle
\flushbottom

\section{Introduction}
\label{Introduction}
Nucleus is a many-body system in which nucleons are bound together by the strong nucleon-nucleon interaction against the repulsive Coulomb interaction. Some traditional low-energy nuclear models, where the most obvious degrees of freedom are protons and neutrons such as the semi-empirical mass formula, shell model and the Hartree-Fock approach, have been proposed and can successfully predict many nuclear properties \cite{LunneyRMP2003}. In contrast with the traditional models, the Skyrme model,  a nonlinear theory of pions originally written down by Skyrme in the 1950s \cite{SkyrmePRSL1954, SkyrmePRSLA1961} is a low energy approximation of quantum chromodynamics (QCD), appropriate when modeling physics in which the quark structure is negligible \cite{WittenNPB1983}. A nucleus with mass number $A$ arises in the Skyrme model as a topological soliton solution, known as a Skyrmion, with the topological charge $B=A$. The Skyrme model has been successful in describing light nuclei (B$\leq$16) \cite{AdkinsNPB1983, AdkinsNPB1984, BraatenPRL1986, BraatenPRD1988, BattyePRC2009, MankoPRC2007, ChrisPRC2017}.

Skyrmions with massive pions exhibit a cluster structure in which low charge baryons serve as subunits of larger baryons \cite{BattyeNPB2005, BattyePRC2006, BattyePRSA2006}. This is more compatible with real nuclei than the Skyrmions with massless pions which have a Fullerene-type structure \cite{BattyePRL1997, BattyePRL2001}. In fact, alpha-particle clusters arise in many theoretical and experimental studies of structure in light nuclei \cite{WheelerPR1937, FreerRPP2007, RenPRL2013}, such as in shell \cite{PerringPPSLSA1956} and lattice $ab$ $initio$ calculations \cite{EpelbaumPRL2011, EpelbaumPRL2014}. The cubic B=4 Skyrmion serves as the alpha-particle in the Skyrme model and it can be used to construct larger Skyrmions \cite{BattyeNPB2005} up to $B=108$ \cite{FeistPRC2013}. After quantization, the excitation spectra of Skyrmions may be compared to those of light nuclei \cite{AdkinsNPB1983, AdkinsNPB1984, BraatenPRL1986, BraatenPRD1988, BattyePRC2009, MankoPRC2007, ChrisPRC2017}. There are some successes, especially for Carbon-12. A $D_{3h}$-symmetric triangle-like Skyrmion describes the ground state and its rotational band while a $D_{4h}$-symmetric chain-like Skyrmion describes the excitations and the root-mean-square charge radius of the Hoyle state successfully \cite{LauPRL2014}.

A persistent problem in the Skyrme model is that the binding energies of the Skyrmions are one order of magnitude larger than that of the experimental results. Motivated by this puzzle, a series of alterations to the classical Lagrangian have been proposed. For example, one may include vector mesons in the Lagrangian or add a new potential energy term which unbinds the Skyrmion \cite{GillardNPB2015, BjarkePRD2016, BjarkePRD2018}.
A Bogomol'nyi-Prasad-Sommerfield (BPS) Skyrme model has solutions which saturate the Bogomol'nyi bound and thus they fix the binding energy problem in theory \cite{AdamPLB2010, AdamPRL2013}. However, the actual Skyrmions remain unknown and hence it is very difficult to make reliable comparisons with experimental data. In addition, many of these models require fine-tuning of parameters and hence are unnatural. The latest development is to include rho mesons. In this extended model, Skyrmions show a clear cluster structure and have low binding energy without having to fine-tune parameters \cite{NayaPRL2018}.

In the Skyrme model, the classical mass of the Skyrmion arises from the contribution of the nuclear strong force. The Coulomb force, an essential ingredient in nuclear systems which naturally unbinds nuclei, is usually ignored in the Skyrme model. Hence one should quantitatively and qualitatively estimate the Coulomb energy contribution to Skyrmions. It is important to check that the Skyrme model correctly reproduces the well known theoretical results for the Coulomb energy of nuclei. In addition, the Coulomb energy could be an important aspect to account for when calculating binding energies of Skyrmions in the standard or loosely bound Skyrme models.

In this paper, we simply introduce the Skyrme model in Section \ref{Skyrme-model}. In Section \ref{Calculation}, we bypass the traditional Coulomb energy calculation, which is usually extracted from proton-proton interactions, by calculating the Coulomb energy using the full numerical Skyrmion solutions and then by using an alpha-like particle approximation where we replace the cubic $B=4$ Skyrmion with an effective alpha-like particle. This approximation works for Skyrmions with baryon number $4N$. In Section \ref{Results}, we compare the results from the Skyrme model with the semi-empirical mass formula and investigate the Coulomb effect on nuclear binding energy. Finally, a summary and outlook is given in Section \ref{Summary}.

\section{The Skyrme model}\label{Skyrme-model}
In the Skyrme model, an isotriplet of pion fields $\pi(x,t)$ and auxillary field $\sigma(x,t)$ are combined into an SU(2)-valued Skyrme field, $U(x,t)=\sigma(x,t)\mathbb{I2}+i\pi(x,t)\cdot \bf{\tau}$. To ensure that $U$ lies in SU(2) one demands that $\sigma^2+\pi\cdot\pi=1$. In physical units, the Lagrangian of the Skyrme model with a pion mass term is \cite{BattyeNPB2005}
\begin{equation} \label{Lagrangian}
 \mathcal{L}=\frac{f_\pi^2}{16}\mathrm{Tr}\partial_\mu U\partial^\mu U^\dag +\frac{1}{32e^2}\mathrm{Tr}([\partial_\mu U U^\dag,\partial_\nu U U^\dag])^2\\
+\frac{1}{8}m_\pi^2f_\pi^2\mathrm{Tr}(U-\mathbb{I}_2).
\end{equation}
In total there are three parameters: $f_\pi$, $e$ and $m_\pi$ are the pion decay constant, a dimensionless coupling parameter and the pion mass respectively. For a static finite energy configuration, $U$ must lie in the vacuum at $|x|\rightarrow \infty$. This provides a one-point compactification of space, $\mathbb{R}^3\bigcup\{\infty\}\simeq S^3$, meaning that the Skyrme field is a map from $S^3$ to the target space, $SU(2)\equiv S^3$. Maps between spheres have a non-trivial homotopy class and can thus be characterized by an integer $B$, identified with the baryon number, which can be written as
\begin{equation}\label{baryon-density}
B=\frac{-1}{24\pi^2}\int \epsilon_{ijk}\mathrm{Tr}[(\partial_iU)U^\dag(\partial_jU)U^\dag(\partial_kU)U^\dag]d^2\bf{x}.
\end{equation}
A Skyrmion is a minimizer of the potential energy of \eqref{Lagrangian} and is labeled by the baryon number $B$. It is then identified with the nucleus with mass number $A=B$.

\section{Calculation}\label{Calculation}
For non-relativistic theories, the Coulomb energy is given by
\begin{equation}\label{classical-Coulomb-expression}
E_C=
\frac{1}{2\varepsilon_0}\int\int d^3rd^3r^\prime\frac{\rho(\vec{r})\rho(\vec{r^\prime})}{4\pi|\vec{r}-\vec{r^\prime}|},
\end{equation}
where $\varepsilon_0=\frac{1}{e}8.8542\times10^{-21}$ $\frac{1}{\mathrm{MeV}\cdot\mathrm{fm}}$ is the permittivity of free space, $e=1.60218\times10^{-19}$ C is the electric charge and $\rho$ denotes the charge density. Unless the charge density is a simple function, it is a challenge to directly calculate Eq. \eqref{classical-Coulomb-expression} due to the divergent behavior at $r=r^\prime$; but one can convert the integral into a series expansion where $\rho(\vec{r})$ is expanded into spherical harmonics $Y_{lm}(\theta, \phi)$, i.e. $\rho(\vec{r})=\sum_{l,m}\rho_{lm}(r)Y^\ast_{lm}(\theta,\phi)$. Following the approach given in \cite{CarlsonAJP1963}, the quantities $Q_{lm}(r)$ can be defined as
\begin{equation}
Q_{lm}(r)=\int_0^r dr^\prime r^{\prime l+2}\rho_{lm}(r^\prime),\label{Qlm}
\end{equation}
which, at large $r$, approach the multipole moments of a Skyrmion charge density. By using  Parseval's theorem for Hankel transformations \cite{CarlsonAJP1963}, the Coulomb energy associated with the configuration can be written as
\begin{equation}
E_C=\sum_{l=0}^{\infty}\sum_{m=-l}^{m=l}U_{lm},
\label{classical-Coulomb-expansion}
\end{equation}
where
\begin{equation}
U_{lm}=\frac{1}{2\varepsilon_0}\int_0^\infty drr^{-2l-2}|Q_{lm}(r)|^2. \label{Ulm}
\end{equation}
Here, the Coulomb energy is independent of the conversion between physical energy units and Skyrme units.

\subsection{$B=1$ Skyrmion}
Since a single Skyrmion is a extended object, as opposed to a point particle in traditional nuclear models, we first calculate the Coulomb energy for $B=1$ Skyrmion, i.e. proton and neutron, which will be used later for evaluating the effect of Coulomb energy on nuclear binding energy. The $B=1$ Skyrmion is spherically symmetric and hence the Skyrme field can be written as $U^\mathrm{H}=\mathbb{I}_2\cos f(r)+i\hat{x}^a\tau^a\sin f(r)$ \cite{AdkinsNPB1983}, where $f(r)$ is a function of $r$, $\hat{x}^a=x^a/r$ is the spatial unit 3-vector. The corresponding radial charge density for the nucleon is $\hat{\rho}=\frac{1}{2}\mathcal{B}^0+\mathcal{I}^3$. $\mathcal{B}^0$ is the baryon charge density [see the integrand of Eq. \eqref{baryon-density}]. $\mathcal{I}^3$ is the third component of the isospin density that comes from the vectorial Noether current and satifies (here $f_r\equiv \partial_r f$)
\begin{equation}
\int d\Omega\mathcal{I}^3=\frac{2\sin^2f+\frac{2}{r^2}\sin^2f(\sin^2f+r^2f_r^2)}{2\int dr(2r^2\sin^2f+2\sin^2f(\sin^2f+r^2f_r^2)},
\end{equation}
where $\Omega$ are the angular spatial coordinates. The symmetry of the charge density enforces constraints on the allowed values of $U_{lm}$. In the case of spherical symmetry, only $U_{00}$ is non-zero. The resulting Coulomb energies for the proton and neutron can be evaluated once one has numerically calculated the profile function $f(r)$. We find that $E_C^\text{p}=0.116$ MeV and $E_C^\text{n}=0.28\times10^{-3}$ MeV.
%
\subsection{Alpha-like particle approximation}
The nucleons' individual identities disappear when brought close together in the Skyrme model. In particular the $B=4$ cubic Skyrmion, which serves as the building block for larger Skyrmions, does not obviously contain four distinct nucleons. Hence we introduce an effective alpha-like particle approximation (APA) to describe larger Skyrmions by treating each $B=4$ cubic cluster as an alpha-like particle with charge $Q_{\text{eff}}$.
Before proceeding, we first calculate the Coulomb energy for $B=4, 8$ Skyrmions by using the numerically generated multi-Skyrmions. In the $B=4$ sector, the Skyrmion has cubic symmetry and a spin-0, isospin-0 ground state. Thus the charge density of the ground state is simply $\hat{\rho}=\frac{1}{2}\mathcal{B}^0$. The $B=8$ Skyrmion is approximately described as two $B=4$ Skyrmions glued together. This has the same quantum ground state as the $B=4$ Skyrmion. The Coulomb energies of the $B=4$ and $8$ Skyrmions, calculated using the numerically generated charge density, are [MeV]
\begin{equation}
E_C^4\approx U_{00}+U_{40}+U_{4-4}+U_{44}=1.95,\label{4+8-Coulomb-results}
\end{equation}
\begin{equation}
E_C^8\approx U_{00}+U_{20}+U_{40}+U_{4-4}+U_{44}=5.45.\label{4+8-Coulomb-results}
\end{equation}
The possible contributing terms $U_{lm}$ for the aforementioned Skyrmions get smaller as $l$ increases. Hence we only include up to $l=4$.

In the alpha-like particle approximation framework, the charge densities of the $B=4$ and $8$ Skyrmions can be approximately constructed using delta functions: these are $\rho_4(r)=Q_{\text{eff}}\delta(r)$ and $\rho_8(r)=Q_{\text{eff}}[\delta(r-r_1)+\delta(r-r_2)]$ respectively. $r_1$ and $r_2$ mark the centers of the two cubes which compose the $B=8$ Skyrmion. Here, we assume that all cubes have the same charge as they describe identical alpha-particles. Inserting $\rho_4(r)$ and $\rho_8(r)$ into Eq. \eqref{classical-Coulomb-expression}, we find that $E_C^\mathrm{8}=2E_C^\mathrm{4}+\frac{2Q_{\text{eff}}^2}{2\varepsilon_0\times4\pi}\frac{1}{|r_1-r_2|} .$ Note that in this letter we have taken the Skyrme length unit to be 1 fm, giving the radius of the $B=4$ Skyrmion as 1.2 fm which is close to the experimental value. We then take $r_1$ and $r_2$ from the numerically generated Skyrmion and find that $|r_1-r_2|\approx2.4$ fm. The effective alpha-like particle charge is then fixed as $Q_{\text{eff}}=1.614$ C. Naively, one would expect that $Q_{\text{eff}}=2$ but this fails to account for the fact that the $B=4$ Skyrmion is an extended object.

\begin{figure}[htbp]
\centering
\includegraphics[width=0.9\textwidth]{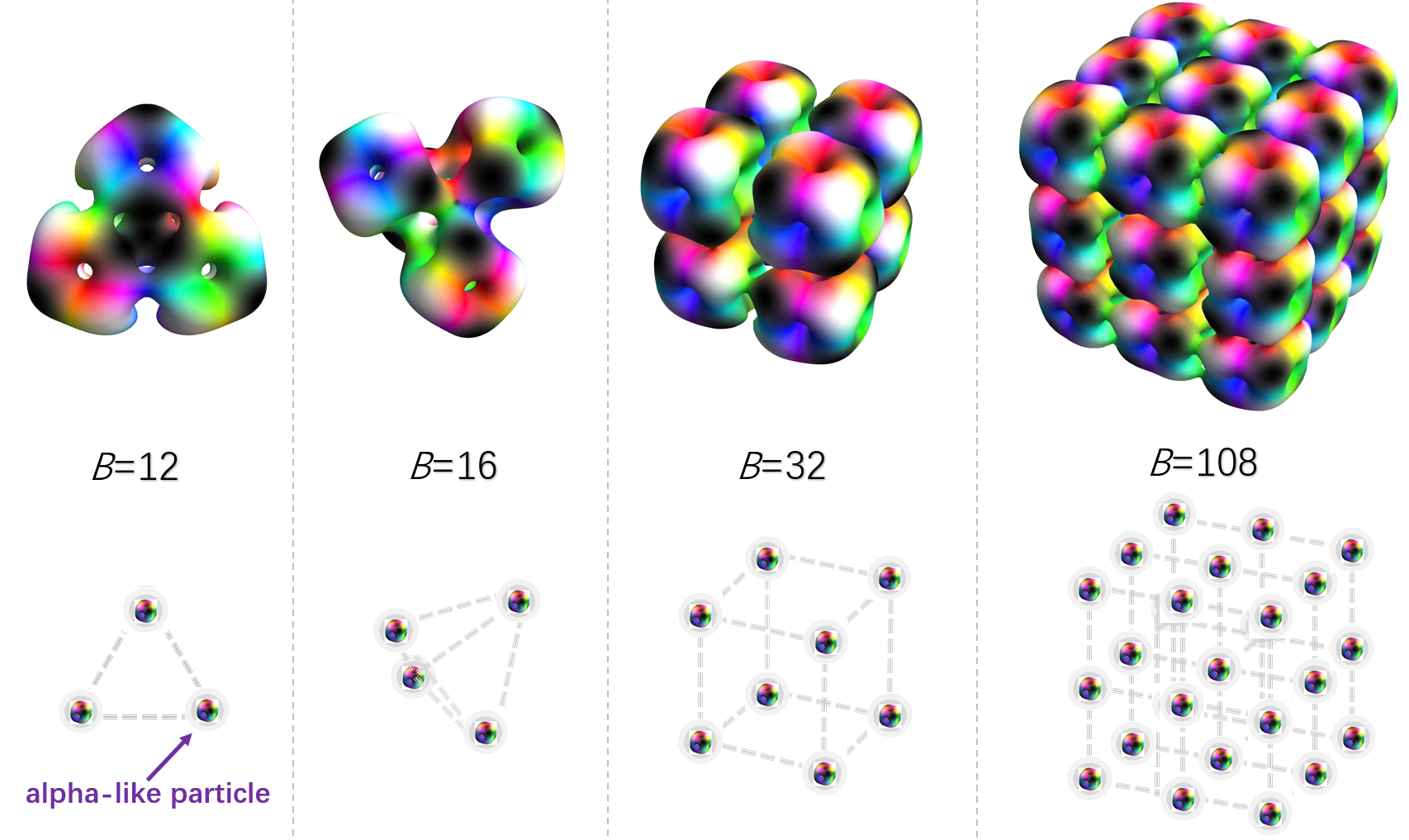}
\caption{A surface of constant baryon density for the $B=12, 16, 32, 108$ Skyrmion (upper row) -- different colors indicate the direction of the pion field. The corresponding simplified sub-figures in the alpha-like particle framework (bottom row).
}\label{configuration}
\end{figure}

\subsection{Multi-Skyrmions}
Secondly, we generate the multi-Skyrmion numerical configurations using a gradient flow method and plot these configurations for $B=12,~16,~32$ and 108 Skyrmions, which have a cluster structure in the standard Skyrme model, in Fig. \ref{configuration}. The coloring represents the direction of the pion field on the baryon density contours. One finds that the $B=12$ Skyrmion has $D_{3h}$ symmetry: the geometrical configuration is an equilateral triangle. The $B=16$ Skyrmion has tetrahedral symmetry while the $B=32$ solution with $O_{h}$ symmetry is composed of eight $B=4$ cubes. As in the $B=32$ case, the $B=108$ Skyrmion is a cube but is made from $27$ sub-cubes. Details of these ground-state configurations can be found in \cite{LauPRL2014, ChrisPRC2017, FeistPRC2013}. The corresponding simplified sub-figures using the alpha-like particle approximation are plotted in the bottom panels in Fig. \ref{configuration}. Balls represent the effective alpha-like particles---the geometrical centers of the composing cubes. Taking account of the electrostatic energies between alpha-like particles, a Coulomb energy expression for Skyrmions in the APA framework can be written as the sum of two parts. The first part denotes the contribution from the cube interacting with itself and the second part represents the interaction between a given cube and its neighbours. The detailed expression is
\begin{equation}
E_C^B=\frac{B}{4}E_C^4+D\sum_{0}^{N}\frac{a_n}{\sqrt{n}},~~~~~~N=\left(\frac{B}{4}\right)^2-\frac{B}{4},\label{general-term}
\end{equation}
where $\frac{B}{4}$ is the number of composing cubes and $D=\frac{Q_{\text{eff}}^2}{2\varepsilon_0 \times4\pi}\frac{1}{2.4}\approx0.781$ MeV. $a_n$ is the number of pairs of clusters which are $\sqrt{n}$ distance apart. Detailed values of $a_n$ for the $B=12, 16, 32$ and $108$ Skyrmions are tabulated in Table \ref{terms121632108}.
\begin{table}[htbp]
\caption{The corresponding parameters in Eq. \eqref{general-term} for $B=12,~16,~32,~108$ Skyrmions. A more detailed interpretation of parameters can be found in the text.} \label{SE}
\begin{center}
\begin{tabular}{ccccccccccccc}
  \hline\hline
 Skyrmions&$a_1$ &$a_2$&$a_3$&$a_4$&$a_5$&$a_6$& $a_7$&$a_8$&$a_9$&$a_{10}$& $a_{11}$&$a_{12}$\\ \hline
$B=12    $    &~6  &--&--&--&--&--&--&--&--&--&--&--\\
$B=16   $     &~12 &--&--&--&--&--&--&--&--&--&--&--\\
$B=32  $      &~24 &24&8 &--&--&--&--&--&--&--&--&--\\
$B=108$       &~108&144&64 &54 &144&96 &--&36 &48 &--&--&8\\
\hline \label{terms121632108}
\end{tabular}
\end{center}
\end{table}
The corresponding Coulomb energies are [MeV]
\begin{equation}
E_C^{12}=10.52,  ~~~~~~E_C^{16}=17.15,~~~~~~~E_C^{32}=51.18,~~~~~~E_C^{108}=371.64. \label{multi-Skyrmion}
\end{equation}

As a check, one can use the full numerical results to compare with the results of the alpha-like particle approximation. We then calculate the $U_{lm}$s by using numerical method (NM) for each Skyrmion and, using these, the total Coulomb energy. The results from this numerical method ($E_C^\mathrm{NM}$) are tabulated in Table \ref{contributions-1-4-8}.

\begin{table}[htbp]
\caption{The values of the $U_{lm}$s for $B=12, 16, 32, 108$ Skyrmions using full numerical method (NM).}
\begin{center}
\begin{tabular}{lcccc}
  \hline\hline
~~~Energy~~~ &\multicolumn{4}{c}{Skyrmions}\\ \cline{2-5}
~~~~[MeV]&~~~$B=12$~~&~~~$B=16$~~&~~~$B=32$~~&~~~$B=108$~~\\ \hline
$U_{00}$          &10.22 &15.43 &48.90&342.97\\
$U_{20}$          &0.33 &-- &--&--\\
$U_{32}~(U_{3-2})$ &-- &0.31 &--&--\\
$U_{33}~(U_{3-3})$ &0.07 &-- &--&--\\
$U_{40}$          &-- &0.05 &0.46&2.33\\
$U_{44}~(U_{4-4})$ &-- &0.02 &0.16&0.84\\ \hline
$E_C^\mathrm{NM}$ &10.72 &16.14 &49.68&347.00\\\hline
\label{contributions-1-4-8}
\end{tabular}
\end{center}
\end{table}

One finds that the values from the alpha-like particle approximation [Eq. \eqref{multi-Skyrmion}] are close to the results of the full numerical method $E_C^\mathrm{NM}$. We therefore verify that the APA framework is an good approximation when dealing with cluster configurations. When one tries to include isospin or go to even larger Skyrmions, it is difficult to use the full numerical method to calculate the Coulomb energy. Thus the alpha-like particle framework may help one to do these more difficult calculations.
\section{Results}\label{Results}
\subsection{Comparison to mass formula}
The semi-empirical mass formula (MF) describes nuclear masses across the nuclear chart, based on empirical and theoretical considerations. The classical mass of a Skyrmion in the Skyrme model roughly plays the same role as the volume and surface terms in MF, which arise from the strong interaction. The Coulomb energy is $E_C^\mathrm{MF}=0.625\frac{Z^2}{A^{\frac{1}{3}}}$ \cite{NingWangPRC2014}.
We focus on isospin-0 nuclei and so for the nuclei in this letter $Z=\frac{B}{2}$. The Coulomb term can then be rewritten as
\begin{equation}
E_C^\mathrm{MF}=0.156 B^\frac{5}{3}. \label{mass-Coulomb-expression-B}
\end{equation}
If the Skyrme model hopes to be a reasonable description of nuclei, it should be able to reproduce equation (\ref{mass-Coulomb-expression-B}).

\begin{figure}[!]
\begin{center}
\includegraphics[scale=0.5]{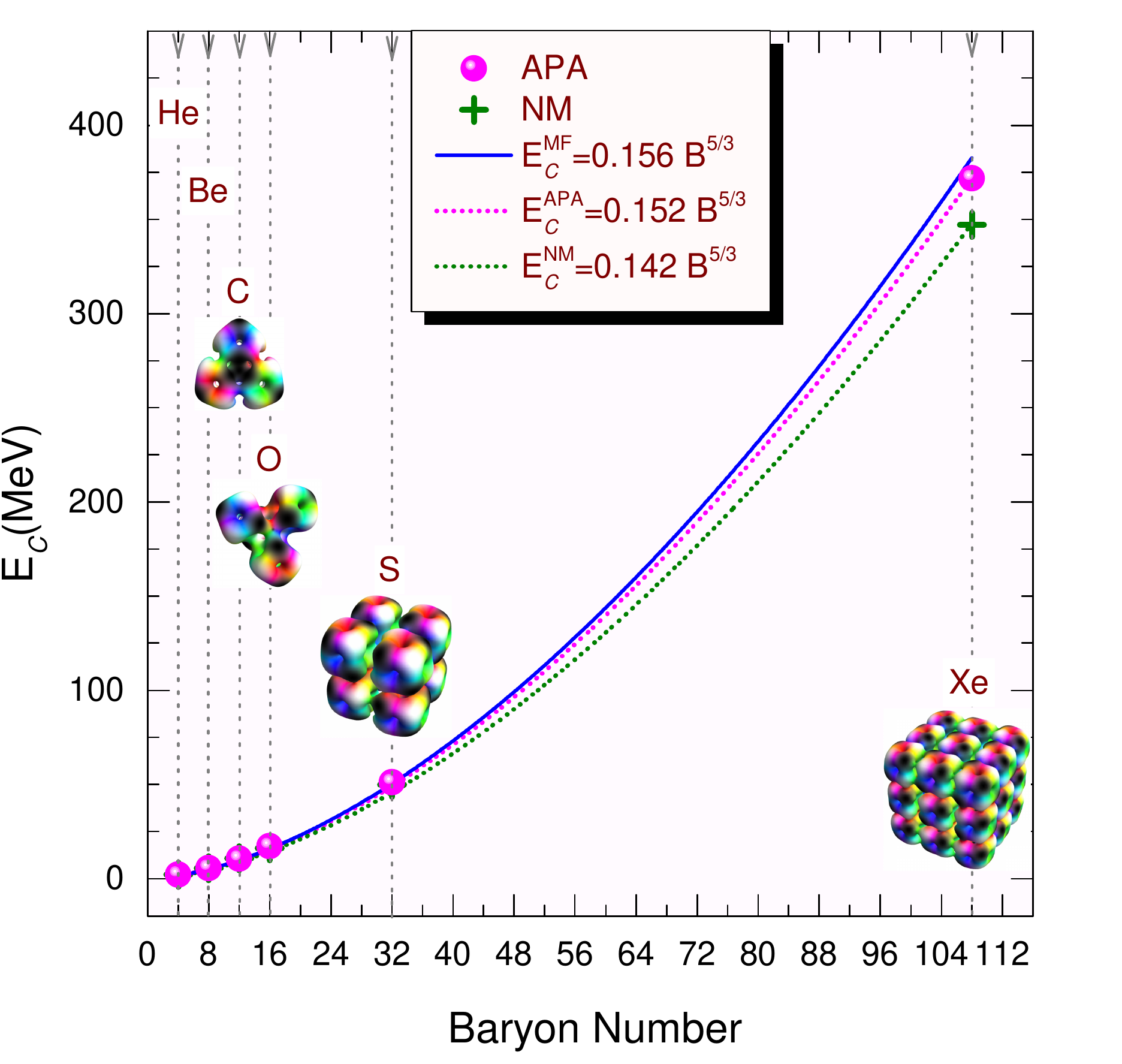}
\caption{Two groups of Coulomb energies for $B=4,~8,~12,~16,~32, ~108$ Skyrmion, APA--magenta spheres, NM--olive crosses. The Coulomb term in MF, two fitted curves---APA as magenta dashed line and NM as olive line, in various of the baryon numbers.
} \label{Coulomb-mass-Skyrme}
\end{center}
\end{figure}

We plot the Coulomb energy of the $B=4N$ Skyrmions from the alpha-like particle approximation, the numerically generated Skyrmions and the expression in the mass formula [Eq. \eqref{mass-Coulomb-expression-B}] in Figure \ref{Coulomb-mass-Skyrme}. One can see that the results from the Skyrme model are in good agreement with the MF for light nuclei with $B<32$. A larger discrepancy appears in the heavy region, though the $B=108$ Skyrmion's Coulomb energy is still within $10\%$ of the value from the MF. Furthermore, we can obtain approximate values of the Coulomb energy in the Skyrme model for any $B$ by fitting our results to a curve of the form $p B^\frac{5}{3}$. We find that
\begin{equation}
E_C^\mathrm{APA}=0.152B^{\frac{5}{3}},\label{Coulomb-epxression-skyrme}
\end{equation}
and these curves can be found in Fig. \ref{Coulomb-mass-Skyrme}. The parameter $0.152$ is close to $0.156$ from Eq. \eqref{mass-Coulomb-expression-B}. Hence we have verified that the Skyrme model correctly reproduces the well known results for the Coulomb energy of nuclei. In addition, the Coulomb energy calculated by treating the $B=4$ Skyrmion as an alpha-like particle building block fits the MF curve well. This suggests that one can describe these nuclei using only alpha-particle degrees of freedom, within the Skyrme model. This agrees with the proposals in other theoretical models in Ref. \cite{ElhatisariPRL2016, PouliotPRC1993, MarquezPRC1983}. One can further explore the alpha-alpha interaction and Coulomb interaction in Skyrmion.
The Coulomb energy for other unknown Skyrmions can be extrapolated by using the expression \eqref{Coulomb-epxression-skyrme}. The APA approximation could provide a good starting point for estimating the Coulomb energy of other Skyrmions, such as those with  $B=4N+n$ \cite{ChrisPRD2018}, lightly/loosely bound Skyrmions \cite{GillardNPB2015, BjarkePRD2016, BjarkePRD2018} and those which are coupled to the rho meson \cite{NayaPRL2018}.

\subsection{Coulomb effect on the nuclear binding energy}
Finally, we calculate the contribution of the Coulomb energy on the nuclear binding energy. The nuclear binding energy is defined as
\begin{equation}
B(Z,N)=ZE^\mathrm{p}+NE^\mathrm{n}-E^B(Z,N) .
 \end{equation}
where $Z$ and $N$ are the proton and neutron numbers and $E^\mathrm{p}$, $E^\mathrm{n}$ and $E^B(Z,N)$ represent the total energy of the proton, neutron and given nucleus respectively. Taking into account the static energy of a Skyrmion $\mathcal{M}_0^{B}$, calculated by solving the static equations of motion, and the Coulomb energy calculated using the alpha-like particle approximation, the total energy of a Skyrmion is $E^B(Z,N)=\mathcal{M}_0^{B}+E_C^\mathrm{APA}$. This neglects significant quantum corrections.

In Table \ref{contribution-to-binding}, we tabulate the decreased values of the Skyrmion binding energy due to the Coulomb energy correction, which we call $\Delta_\mathrm{B}$. For Skyrmions labeled by $^*$ ($B=$20, 24, 28, 40, 100, 104) we take the Skyrmion configuration to be unknown and the corresponding Coulomb energies and decreased energies are extrapolated using the fitting curve $E_C^\mathrm{APA}=0.152B^{\frac{5}{3}}$. One can see that the Coulomb energy has a small effect on the binding energy for light Skyrmions. In the region of heavy nuclei, the Coulomb energy quickly increases with baryon number. This trend is consistent with the behavior of Coulomb energy in the traditional mass formula. For the $B=108$ Skyrmion, $\mathcal{M}_0^{108}\approx 135.47$ and $\mathcal{M}_0^{1}\approx 1.465$ (in Skyrme units/ $12\pi^2$), the total binding energy is reduced by nearly 2.4$\%$ by incorporating the Coulomb energy correction. Although the Coulomb effect does unbind the Skyrmion, it is nowhere near enough to solve the binding energy problem. However, it could be important for solutions in the loosely bound Skyrme models. For example, the experimental binding energy of Helium-4 is 28.296 MeV. The $B=4$ Skyrmion has a classical binding energy of around 400 MeV. The Coulomb energy decreases this by 1.72 MeV. Clearly it is not enough. However, the classical binding energy of the $B=4$ Skyrmion in some loosely bound models is only 8 MeV \cite{BjarkePRD2018}. When the Coulomb energy term is involved, the total classical binding energy could be reduced by nearly 25$\%$. Thus the structure of the loosely bound Skyrmions could be changed. Hence the Coulomb energy plays an important role in loosely bound models.

\begin{table}[htbp]
\caption{
The Coulomb energies and the decreased values of Skyrmion binding energy [MeV]. Baryons labeled by the sign $^*$ have had their Coulomb energies calculated using the fitting curve $E_C^\mathrm{APA}=0.152B^{\frac{5}{3}}$.}
\begin{center}
\begin{tabular}{ccccccccccccc}
\hline\hline
 &$\vdots $&\multicolumn{11}{c}{APA}\\ \cline{3-13}
$B$&$\vdots $&4&$\vdots $&8&$\vdots $&12&$\vdots $&16&$\vdots $&32&$\vdots $&108\\
$E_C$&$\vdots $&1.95&$\vdots $&5.45&$\vdots $&10.52&$\vdots $&17.16&$\vdots $&51.18&$\vdots $&371.64\\
$\Delta_\mathrm{B}$&$\vdots $&1.72&$\vdots $&4.98&$\vdots $&9.82&$\vdots $&16.23&$\vdots $&49.32&$\vdots $&365.36\\
 &$\vdots $&\multicolumn{11}{c}{$E_C^\mathrm{APA}=0.152B^{\frac{5}{3}}$}\\   \cline{3-13}
$B$$^*$&$\vdots $&20$^*$&$\vdots $&24$^*$&$\vdots $&28$^*$&$\vdots $&40$^*$&$\vdots $&100$^*$&$\vdots $&104$^*$\\
$E_C$&$\vdots $&22.38&$\vdots $&30.33&$\vdots$& 39.22&$\vdots $&71.07&$\vdots $&327.26&$\vdots $&349.37\\
$\Delta_\mathrm{B}$&$\vdots $&21.22&$\vdots $&28.94&$\vdots $&39.59&$\vdots $&68.74&$\vdots $&321.45&$\vdots $&343.32\\  \hline \label{contribution-to-binding}
\end{tabular}
\end{center}
\end{table}

\section{Summary and outlook}\label{Summary}
To summarize, we have done a quantitative evaluation of the Coulomb effect on Skyrmions with baryon number $4N$ by using an effective alpha-like particle approximation as well as the full numerical Skyrmion solutions. In the former, we treat $B=4$ Skyrmions---subunits used to construct larger Skyrmions---as alpha-like particles. We have shown that the Coulomb energy from the alpha-like particle approximation is in good agreement with the real model by comparing its results to the full numerical calculations. Further, the results are in broad agreement with the phenomenological Coulomb energy contribution to the nuclear mass formula. Thus we can summarize that (i) the Skyrme model accurately reproduces the correct Coulomb energy behavior; (ii) The Coulomb correction decreases the total binding energy of Skyrmions by a small amount. However, it could reduce the binding energy of the loosely bound Skyrmions by around $25\%$ and thus change their structures. Further, one can extrapolate the Coulomb energy for other unknown Skyrmions and hence its effect on their binding energy. Using the alpha-particle inspired, alpha-like particle approximation one could now attempt to calculate the Coulomb energies for more difficult configurations (like $B=4N + n$ Skyrmions) or in modified Skyrme models such as the lightly/loosely bound models or ones which include the rho meson.

Additionally, this work suggests that the alpha-particle can act as the most important degree of freedom in Skyrme model. This provides a
new window on studying the effect of alpha-alpha interaction on nuclear properties and the cluster structure of Skyrmion, as well as the breakup mechanisms of larger Skyrmion divides into subunits.



\acknowledgments
N. Ma thanks to P. H. C. Lau, Rob. Pisarski and R Wang for useful discussions. C. J. Halcrow is supported by The Leverhulme Trust as an Early Careers Fellow. N. Ma is funded by the National Natural Science Foundation of China (Grant No. 11675223). Partly supported by National Natural Science Foundation of China Grant No. 11675066, by the Fundamental Research Funds for the Central Universities Grant No. lzujbky-2017-ot04 and Feitian Scholar Project of Gansu province.

\end{document}